\documentclass[pre,aps,amsmath,amssymb,superscriptaddress]{revtex4-1}
\usepackage{mathrsfs}
\usepackage{amssymb}
\usepackage{graphicx}
\usepackage{amsmath}
\usepackage{graphicx}
\usepackage{dcolumn}
\usepackage{bm}
\newcommand{\bra}[1]{\left(#1\right)}
\newcommand{\brb}[1]{\left[#1\right]}
\newcommand{\brc}[1]{\left<#1\right>}

\newcommand{\fw}{0.5}

\begin{document}
\title{Inference of kinetic Ising model on sparse graphs}
\author{Pan Zhang }

\affiliation{Politecnico di Torino, C.so Duca degli Abruzzi 24, 10129 Torino, Italy}
\begin{abstract}
  Based on dynamical cavity method, we propose an approach to the inference of kinetic Ising model, which asks to reconstruct couplings 
  and external fields from given time-dependent output of original system. Our approach gives an exact result on tree graphs and a good 
  approximation on sparse graphs, it can be seen as an extension of Belief Propagation 
  inference of static Ising model to kinetic Ising model. While existing mean field 
  methods to the kinetic Ising inference e.g., na\" ive mean-field, TAP equation and simply mean-field, use approximations which calculate 
  magnetizations and correlations at time $t$ from statistics of data at time $t-1$, dynamical cavity method can use statistics of data at times earlier than 
  $t-1$ to capture more correlations at different time steps.
Extensive numerical experiments show that our inference method is superior to existing mean-field approaches on diluted networks.
\end{abstract}
\maketitle

\section{introduction}
Inference problem, which aims at learning internal 
structure of complex systems from empirical output data, has been studied for a long time. 
Among many models, inverse Ising model, which is a basic pair-wise model in physics, has been widely studied for inference problem in many fields.

A lot of study\cite{Cocco_etal_PNAS_206_14058_2009,Weigt_etal_PNAS_2009,Ackley1985147,Roudi_etal_PRE_79_051915} has been carried out on the (static) inverse Ising problem, which asks to 
infer couplings and external fields of Ising model from configurations sampled from Boltzmann distribution. Efficient algorithms based on mean-field approximations 
and message passing have been developed to address the inference task. Recently it receives special attention in neural network reconstruction of retinas based on 
multi-electrode recordings\cite{Schneidman_etal_2006_nature}, and gene-gene interaction reconstruction\cite{Weigt_etal_PNAS_2009}.
In static inverse Ising model, empirical data are $M$ equilibrium configurations 
$\{{\boldsymbol\sigma}_1,{\boldsymbol\sigma}_2,\cdots,{\boldsymbol\sigma}_M\}$ that sampled from 
Boltzmann distribution $\mathcal{P}=\frac{1}{Z}e^{-H({\boldsymbol\sigma})}$, where energy ${H({\boldsymbol\sigma})}$ is a function of couplings and external fields. When $M$ is very 
large, posterior distribution of $H$ peaks on the maximum point so that one can find the couplings by maximizing the log-likelihood $\mathcal{L}=-\brc{H}-\log Z$, where 
$\brc{H}$ denotes averaged energy. Thanks to 
the convexity of log-likelihood, the exact method, so called Boltzmann machine, computes exactly the correlations and magnetizations and match them with empirical ones. 
However, computing magnetizations and correlations exactly for large systems is unreachable, so many efforts have been paid to study approximate approaches 
based on different approximations, including na\" ive mean-field approximation, TAP equation, small correlation expansion and message passing schemes.

Boltzmann distribution is a strong restriction to system to be studied. Many real systems like biological systems, 
especially with time-dependent external stimuli, do not have this good property. \textit{Inverse kinetic Ising model}, which asks to reconstruct couplings and time-dependent external fields
from dynamic output of real system, obviously meets wider needs in real system inference. In inverse kinetic Ising model, exact reconstruction uses all empirical configurations 
in learning and time-complexity is proportional to the number of configurations. When number of configurations is large, one needs efficient approximate methods. 
In last years, Na\" ive Mean-field and TAP approaches have been adapted from static Ising model in case of weak couplings\cite{Roudi_Hertz_PRL_106_048702}. On densely connected fully asymmetric networks, an 
exact reconstruction method is proposed for asymmetric SK model\cite{Mezard_Sakellariou_jstat_2011,Sakellariou_etal_arxiv_2011}. 

However, biological systems are often sparsely connected, so it should be valuable to adapt message passing\cite{Braunstein_etal_PRE_2011} method which work well in sparse-connected static Ising model inference
into kinetic Ising model inference. Another problem of existing mean-field methods is that they use approximations which predict quantities of time $t+1$ based on time $t$, 
we term these approximations 
\textit{one-step approximations}. There are two cases that one-step approximations works well. In first case, experimental data are drawn from stationary state 
where $t\to\infty$ and in second case there is no feedback loops in system e.g. the fully asymmetric model. But in practice it is difficult to ensure these two conditions,
so it should be useful to develop a 
method that use statistics at not only one time step before, but earlier time steps to do the construction. 
In what follows, we show how to use Bethe approximation in inference of kinetic Ising model, in context of dynamic cavity method. This can be seen as an 
extension of Belief-Propagation inference in static inverse Ising problems\cite{Braunstein_etal_PRE_2011}.

The paper is organized as follows. Firstly, in Sec. \ref{sec:model}, we give definition of kinetic Ising model. In Sec. \ref{sec:exact} we introduce exact inference and existing 
mean-field approaches. In Sec. \ref{sec:dc} we discuss dynamical cavity methods and inference of kinetic Ising model. Sec. \ref{sec:result} contains some numerical 
results to compare performance of inference by dynamical cavity method with other mean-field approximations. The last section contains conclusions and some discussions.

\section{Kinetic Ising Model\label{sec:model}}
We consider random graphs with $N$ vertices and average connectivity $C$. In this paper we focus on sparse graphs that $C\ll N$. Connections of graphs are defined 
by matrix $c_{ij}={0,1}$. $c_{ij}=1$ means there is a directed edge from vertex $i$ to 
vertex $j$ and $c_{ij}=0$ means there is no such edge.
If $c_{ij}=c_{ji}=1$, the edge between $i$ and $j$ is symmetric, otherwise it is asymmetric. We set each edge to be 
symmetric with probability $P_s$ and asymmetric with probability $1-P_s$. Obviously, with $P_s=1$, graph is undirected and with $P_s=0$, graph is fully asymmetric. 

On each vertex $i$ there is a spin $\sigma_i=\{+1,-1\}$ associated, and on each edge $j\to i$ there is a coupling $J_{ij}$ associated. If $c_{ji}=0$,  coupling 
$J_{ij}$ is set to $0$ otherwise value of $J_{ij}$ is generated randomly according to Gaussian distribution with zero mean and variance $J/\sqrt{C}$. Dynamics of spins are 
defined by configurations at different discrete time steps from time $0$ to time $T$: ${\underline\sigma(0)},\dots,\underline\sigma(T)$. At each time step, value of spin $i$ is updated
based on configuration of last time according to 
\begin{equation}\label{eq_wi}
  W[{{\sigma}_i\bra{t+1}}|{\boldsymbol\sigma\bra{t}}]=\frac{\exp\left[\beta\sigma_i\bra{t+1}h_i(t)\right]}{2\cosh{\beta h_i(t)}},
\end{equation}
where $\beta$ denotes the inverse temperature and local field at time $t$ is expressed as
$  h_i(t)=\sum_{j\in\partial i}J_{ij}\sigma_j\bra{t}+\theta_i(t),$
with $\theta_i(t)$ denoting external fields acting on spin $i$ at time $t$. We consider in this paper the parallel dynamics, that all spins updated synchronously, evolution of configuration is written as
\begin{equation}\label{eq_wip}
  W[{{\boldsymbol\sigma\bra{t+1}}}|{\boldsymbol\sigma\bra{t}}]=\prod_{i=1}^NW[{{\sigma}_i\bra{t+1}}|{\boldsymbol\sigma\bra{t}}].
\end{equation}
We argue that it would be not difficult to extend our approach from parallel dynamics to
sequential dynamics following what has been done for mean-field methods in \cite{Zeng_etal_PRE_2011}.

Direct problem of kinetic Ising model asks to predict dynamical behavior of spins at an arbitrary time $t$, given couplings, external fields and initial state of network.
Calculating full distribution of system $P({\boldsymbol\sigma(t)})$ is a very difficult task, so most theories focused on 
computing macroscopic observables e.g. magnetization at a time $m(t)=\sum_{\sigma_i(t)}\sigma_i(t)p(\sigma_i(t))$ and correlation at different times 
$c_{ij}(t,t')=\sum_{\sigma_i(t),\sigma_j(t')}\sigma_i(t)\sigma_j(t')p(\sigma_i(t),\sigma_j(t'))$. This direct problem have been studied for long time especially in the 
context of attractor neural networks\cite{Coolen:arxiv:2000b,Hatchett_etal_JPA_37_6201_2004}. Many approaches has been proposed to study ensemble averaged macroscopic quantities, 
e.g. dynamical replica method\cite{Coolen_Sherrington_PRB_1996}, generating functional method\cite{Sommers_PRL_1987} and dynamical cavity method{\cite{Neri_Bolle_JSM_2009}}.

Inverse problem of kinetic Ising model is the problem we would like to study in this paper. Unlike direct problem, instead of couplings and external fields,
experimental output of original model (set of configurations) is given, one is asked to infer couplings and external fields from those experimental data.
The experimental data are obtained by running parallel dynamics of Ising model according to eq.(~\ref{eq_wi}) on graphs for $R$ realizations of time length $T$ paths(trajectories),
denoted by $\{\sigma_i^{r}\bra t\}$ with $r\in [1,R]$ and $t \in [1,T]$. In this paper, we set $T$ to $10$.

\section{Exact reconstruction and mean-field approximations}\label{sec:exact}
Given experimental configurations, probability of observing these samplings as a function of couplings is written as:
\begin{eqnarray}
  P\bra{{\boldsymbol J_{ij}};{\boldsymbol \theta_i(t)}}&=&\prod_{r=1}^R\prod_{t=1}^{T-1}  W[{{\boldsymbol\sigma\bra{t+1}}}|{\boldsymbol\sigma\bra{t}}]\nonumber\\
  &=&\prod_{r=1}^R\prod_{t=1}^{T-1}\prod_{i=1}^N\exp\left [\beta\sigma_i^r\bra{t+1}h_i^r(t)
  - \log 2\cosh\beta h_i^r(t)\right].
\end{eqnarray}
Then, log-likelihood of observing data is defined as:
\begin{eqnarray}
  \mathcal{L}\bra{{\boldsymbol J_{ij}};{\boldsymbol \theta_i(t)}}&=&\log\bra{P\bra{{\boldsymbol J_{ij}};{\boldsymbol \theta_i(t)}}}\nonumber\\
  &=&\sum_{r=1}^R\sum_{t=1}^T\left[\beta\sum_{i=1}^N\sigma_i^r\bra{t+1}\bra{\sum_{j\in\partial i}J_{ij}\sigma^r_j\bra{t}+\theta_i(t)} - \sum_{i=1}^N\log2\cosh\beta\bra{\sum_{j\in\partial i}J_{ij}\sigma^r_j\bra{t}+\theta_i(t)}\right] .
\end{eqnarray}
To find the most-likely couplings and external fields, one needs to maximize the log-likelihood.
At the maximum point of log-likelihood, by setting derivatives of $\mathcal{L}$ with respect to $J_{ij}$ and $\theta_i$ to zero, one has following equations:
\begin{eqnarray}
  \brc{\sigma_i^r\bra {t+1}}_R &=& \brc{\tanh\beta\bra{\sum_{k\in\partial i}J_{ik}\sigma^r_k\bra{t}+\theta_i(t)}}_R.\label{eq:equalm} \\
  \brc{\brc{\sigma_i^r\bra {t+1} \sigma_j^r\bra t}_T}_R &=& \brc{\brc{\tanh\brb{\beta\bra{\sum_{k\in\partial i}J_{ik}\sigma^r_k\bra{t}+\theta_i(t)}}\sigma^r_j\bra{t}}_T}_R\label{eq:equalj} ,
\end{eqnarray}
where $\brc{\cdot}_T$ denotes taking average over time and $\brc{\cdot}_R$ denotes taking average over realizations. 

Let us use $m_i^{data}(t)=\brc{\sigma_i^r\bra {t}}_R$ and $c_{ij}^{data}(t,t-1)=\brc{\brc{\sigma_i^r\bra {t} \sigma_j^r\bra t-1}_T}_R$ 
to denote experimental magnetization at time $t$ and correlation at time $t$ and $t-1$ respectively, 
and use $m_i^J(t)$  and $c_{ij}^J(t+1,t)$ to denote magnetization and correlations predicted by Ising model.
eqs.~(\ref{eq:equalm}),(\ref{eq:equalj}) show that at max-likelihood point, $m_i^J(t+1)$ and $c_{ij}^J(t+1,t)$ are calculated using configurations at time $t$, so we term it 
\textit{configurations based one-step reconstruction}.
The reconstruction can be carried out by using gradient descent learning starting from an initial couplings and fields:
\begin{eqnarray}
  \theta^{}_{i}=\theta^{}_i+\eta (\brc{m^{data}_{i}(t)}_t-\brc{m^J_{i}(t)}_t)\label{eq:grad:m}\\
  J^{}_{ij} = J^{}_{ij}+\eta (\brc{c_{ij}^{data}(t+1,t)}_t-\brc{c^J_{ij}(t+1,t)}_t)\label{eq:grad:c}
\end{eqnarray}
where constant $\eta$ is learning rate. Note that in above scheme, one has to scan all configurations to compute the average on every learning step. It is time consuming and not realistic when amount of experimental data is huge.
To overcome this computational problem, several mean-field approaches, e.g. Na\" ive mean-field, TAP and simply mean-field method, are proposed. They use different approximations to compute 
macroscopic observables, say average magnetizations and correlations, and use those macroscopic observables, instead of all configurations, to do inference.

Na\" ive mean-field method\cite{Roudi_Hertz_PRL_106_048702} applies na\"{i}ve mean-field approximation
\begin{equation} m_i^J(t+1)=\tanh\brb{\beta\sum_{j\in \partial i}J_{ij}m_j^{}(t)}, \end{equation}
and compute one-step delayed correlation as
\begin{eqnarray}
  c_{ij}^J(t+1,t)&=&m_j^{}(t)m_i^J(t+1)+(1-(m_i^J(t+1))^2)\sum_k J_{ik}\brb{c_{kj}^{}(t,t)-m_k^{}(t)m_j^{}(t)}.
\end{eqnarray}
 As an extension of na\" ive mean-field, TAP method takes the expansion on couplings and outperforms na\" ive TAP method in case of weak couplings, the approximation is written as:
 \begin{equation}\label{eq:tapm}
 m_i^J(t+1)=\tanh\brb{\beta\sum_{j\in \partial i}J_{ij}m_j^{}(t) -m^J_i(t+1)\sum_jJ_{ij}^2(1-m_j^{2}(t))},
\end{equation}
and delayed correlation is expressed as
\begin{eqnarray}
  c_{ij}^J(t+1,t)&=&m_j^{}(t)m_i^J(t+1)+
  \brb{1-(m_i^J(t+1))^2}\beta\brb{1-\brb{1-(m_i^J(t+1))^2}\sum_kJ_{ij}^2(1-m_k^{2}(t)}\nonumber\\
  &\times&\sum_k J_{ik}\brb{c_{kj}^{}(t,t)-m_k^{}(t)m_j^{}(t)}.
\end{eqnarray}
Another recently proposed approach, simply Mean-field method\cite{Mezard_Sakellariou_jstat_2011,Sakellariou_etal_arxiv_2011}, which assumes Gaussian distribution of the local field of each spin, computes magnetization as 
\begin{equation} m_i^J(t+1)=\int \frac{dx}{2\pi}e^{x^2/2}\tanh{\beta\bra{\sum_{j\in \partial a}J_{ij}m^{}_j(t)+x\sqrt{\sum_jJ_{ij}^2(1-m_j^{2}(t))}}},\end{equation} and delayed correlation as
\begin{eqnarray}
  c_{ij}^J(t+1,t)&=&m_j^{}(t)m_i^J(t+1)+
  \int \frac{dx}{2\pi}e^{x^2/2}\brb{1-\tanh^2\beta\bra{\sum_{j\in \partial a}J_{ij}m^{}_j(t)+x\sqrt{\sum_jJ_{ij}^2(1-m_j^{2}(t))}}}\nonumber\\
  &\times&\sum_k J_{ik}\brb{c_{kj}^{}(t,t)-m_k^{}(t)m_j^{}(t)}.
\end{eqnarray}
In above listed mean-field approximations, predictions of $m_i^J(t+1)$ and $c_{ij}^J(t+1,t)$ are made by using magnetizations and correlations at time $t$, 
this kind of reconstruction can be termed \textit{statistics based one-step reconstruction}. In some special cases, e.g. fully asymmetric networks, 
observables at time $t$ are sufficient to predict observables at time $t+1$ since effect of feedback loops are not present.
But In general cases, to predict observables at time $t+1$ one needs to use not only observables at time $t$ but also at earlier time steps up to $t=0$.

\section{Dynamical cavity reconstruction}\label{sec:dc}
Dynamical cavity method, originally proposed in \cite{Neri_Bolle_JSM_2009}, together with dynamical replica analysis and generating functional analyses, 
are powerful tools in analysing dynamics of networks. The advantage of dynamical cavity method with respect to other two methods is that it can be applied on single instances to compute
observables for every node.

From eq. ~(\ref{eq_wip}), we can write the probability of observing a path as:
\begin{eqnarray}
  P({\boldsymbol \sigma^{[0,T]}})=\prod_{t=0}^TW[{\boldsymbol \sigma(t+1})|{\boldsymbol \sigma(t)}]P({\boldsymbol \sigma(0)}),
\end{eqnarray}
where $P({\boldsymbol \sigma(0)})$ denotes initial probability distribution of spin configuration at time $0$.
It is difficult to study directly the full path distribution, so dynamical cavity method focuses on marginal probability 
of a path that spin $i$ evolves from time $0$ to time $t$ conditioning on the local fields $\theta_i^{[0,t]}$:
\begin{equation}
  P_{i}({\boldsymbol \sigma_i^{[0,t]}}|\theta_i^{[0,t]})=\sum_{ {\boldsymbol \sigma_{j\neq i}^{[0,T]}}}P(\{{\boldsymbol \sigma_i^{[0,t]}};{\boldsymbol \sigma_j^{[0,T]}}\}|{\boldsymbol \theta^{[0,t]}}).
\end{equation}
By applying Bethe approximation, one can derive following iterative cavity equations (we refer to literature \cite{Neri_Bolle_JSM_2009} for details of derivations):
\begin{eqnarray}\label{eq:cavity}
P_{i\to k}({\boldsymbol \sigma_{i}^{[0,t]}}|{\boldsymbol \theta_i^{[0,t-1]}}+J_{ki}{\boldsymbol \sigma_k^{[0,t-1]}})&=&\sum_{{\boldsymbol \sigma_j^{[0,t-1]}}}
 \prod_{j\in\partial i\backslash k} P_{j\to i}({\boldsymbol \sigma_{j}^{[0,t-1]}}|{\boldsymbol \theta_{j}^{[0,t-1]}}+J_{ji}{\boldsymbol \sigma_i^{[0,t-1]}})\nonumber\\&\cdot&
 \prod_{t'=1}^{t-1}
  \frac{e^{\beta \sigma_i^{t'}[\sum_{j\in\partial i}J_{ij}\sigma_{j}^{t'-1}+
  \theta_i^{t'-1}]}}
  {2\cosh\beta [\sum_{j\in \partial i}J_{ij}\sigma_{j}^{t'-1}
  +\theta_i^{t'-1}]}P_i(\sigma_i^0),
\end{eqnarray}
where $P_{i\to k}({\boldsymbol \sigma_{i}^{[0,t]}}|{\boldsymbol \theta_i^{[0,t-1]}}+J_{ki}{\boldsymbol \sigma_k^{[0,t-1]}})$ is cavity message denoting (cavity) marginal probability 
of a path that spin $i$ evolves from time $0$ to time $t$ with its neighbor spin $k$ removed from the graph. 

Above equation is exact on tree graphs and a good approximation on sparse graphs.
After a fixed point of cavity equation is reached, observables e.g. marginal probabilities, magnetizations and correlations can be computed as functions of cavity messages.
However, solving equation eq.~(\ref{eq:cavity}) is very time-consuming for large $t$ because computational complexity is proportional to $2^{2t}$. So in practice 
one needs approximations to reduce computational complexity. Here we adopt the most simple approximation, one-time approximation\cite{Neri_Bolle_JSM_2009}, 
which is also named time-factorization approximation\cite{Aurell_Mahmoudi_JSM_2011}:
\begin{equation}
P_{i\to k}({\boldsymbol \sigma_{i}^{[0,t]}}|J_{ki}{\boldsymbol \sigma_k^{[0,t-1]}})
=\prod_{t'=0}^{t}P_{i\to k}(\sigma_i^{t'}|J_{ki}\sigma_k^{t'-1}).
\end{equation}
This approximation ignores correlations among different time steps before time $t-2$, and makes summation over variables before time $t-2$ possible.
Then, from eq. (\ref{eq:cavity}), one arrives at 
\begin{eqnarray}
  P_{i\to k}(\sigma_i^k|J_{ki}\sigma_k^{t-1}) &=& \sum_{ \sigma_i^{t-2}} \sum_{{\boldsymbol \sigma_j^{t-1}}}\prod_{j\in\partial i\backslash k} P_{j\to i}(\sigma_j^{t-1}|J_{ji}\sigma_i^{t-2})
  W[\sigma_i^{t}|\sum_{j}J_{ij}\sigma_j^{t-1}]P_i(\sigma_i^{t-2}),
\end{eqnarray}
and expression of magnetizations and delayed correlations can be derived from above equation
\begin{eqnarray}
  m_{i}^J(t+1) &=& \sum_{\sigma_i(t+1)}\sum_{ \sigma_i^{t-1}} \sum_{{\boldsymbol \sigma_j^{t}}}\prod_{j\in\partial i} P_{j\to i}(\sigma_j^{t}|J_{ji}\sigma_i^{t-1})
  W[\sigma_i^{t}|\sum_{j}J_{ij}\sigma_j^{t}]P_i(\sigma_i^{t-1})\sigma_i(t+1)\nonumber\\\\
  c_{ij}^J(t+1,t)&=&\sum_{{\boldsymbol \sigma_k^{t}}} \sum_{{\sigma_j^{t}}} \prod_{k\in \partial i}P_{k\to i}\bra{ {\sigma_k^{t}} |J_{ki}{\sigma_i^{t-1}}} P_{j\to i}\bra{{ \sigma_j^{t}} |J_{ji}{\sigma_i^{t-1}}}\tanh\bra{\beta\sum_{k\in\partial i}J_{ik}\sigma_{k}^{t}}\sigma_{j}^{t}\label{eq:c2}.
\end{eqnarray}
With $m_i^J(t+1)$ and $c_{ij}^J(t+1,t)$ obtained, a standard procedure to perform inference is using gradient descent to learn couplings from differences between experimental and predicted correlations, 
and learn external fields from differences between experimental and predicted correlations.
\begin{eqnarray}
  \theta_i^{Ising}=\theta_i^{Ising}+\eta\Delta m_i\\
  J_{ij}^{Ising}=J_{ij}^{Ising}+\eta\Delta C_{ij},
\end{eqnarray}
where 
\begin{eqnarray}
  \Delta m_i&=&\frac{1}{T}\sum_{t=1}^T\brb{m^{data}_i(t)- m^{J}_i(t)}\\
  \Delta C_{ij}&=&\frac{1}{T-1}\sum_{t=2}^T\brb{C^{data}_{ij}(t,t-1)- C^{J}_{ij}(t,t-1)}.
\end{eqnarray}
Note that there are two ways to do the inference using magnetizations and delayed correlations, one way as used in this paper is to learn couplings by gradient descent,
and the other one is to solve directly the couplings by matching exactly the experimental data, which usually requires inversion of a correlation-based matrix\cite{Roudi_Hertz_PRL_106_048702,Mezard_Sakellariou_jstat_2011}. 
Usually gradient descent learning works slower than matrix inversion method and the convergence of learning method 
may depend on initial couplings and learning rate. However, in matrix inversion method, sometimes it is difficult to find a set of couplings and external fields that
exactly matches the noisy experimental data, while the learning method is able to converge to a fixed point close to the true experimental data (as shown in fig.~\ref{fig:con}).

\section{Performance and comparison with mean-field methods}\label{sec:result}
\begin{figure}
\includegraphics[width=\fw\textwidth]{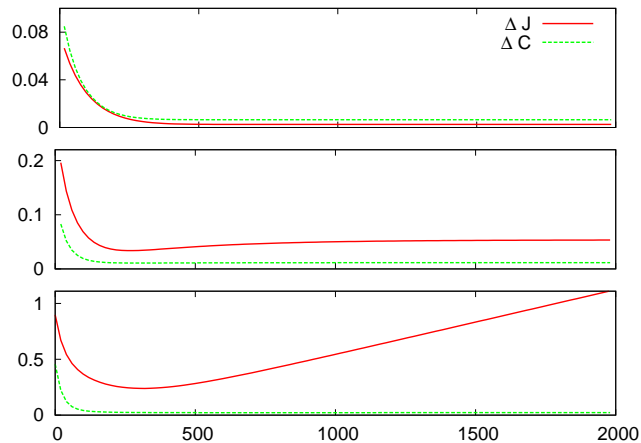}
\caption{ \label{fig:con}
Evolution of inference error $\Delta J$ and difference of one-step delayed correlations of dynamical cavity inference as function of iterating time for networks with $N=100$,
$C=5.0$, $P_s=0$ and different coupling strength. Number of realizations used in experimental data $R$ is $5\times 10^4$. Networks
are fully asymmetric with $J=0.2, 0.9$ and $2.0$ in top, middle and 
bottom panel respectively.}
\end{figure}

In this section we make comparative analysis between dynamical cavity method and mean-field methods on inverse
kinetic Ising problem. Before we compare performance of those methods, 
the first thing we are interested in is the convergence 
properties of gradient descent learning used in our scheme.
In fig.~\ref{fig:con} we plot the evolution of average inference error at each time step
\begin{equation}
\Delta J=\frac{1}{\sqrt{NC}}\sqrt{\sum_{<ij>}(J_{ij}^{Ising}-J_{ij}^{true})^2} \label{eq:dj}
\end{equation}
and average difference of one-step delayed correlation 
\begin{equation}
  \Delta C=\frac{1}{\sqrt{NC}}\sqrt{\sum_{<ij>}(C_{ij}^{J}-C_{ij}^{data})^2}\label{eq:dcj}
\end{equation}
given by dynamical inference in the gradient descent process. In fig.~\ref{fig:con} we recorded 
evolution of $\Delta J$ and $\Delta C$ in three gradient descent process on the same network with different coupling strengths.
We can see that in three panels of fig.~\ref{fig:con}, all correlation differences converge in the learning, 
but evolutions of $\Delta J$ are much different. In case of weak couplings(e.g. $J=0.2$ at left panel),
$\Delta J$ decreases monotonously with difference of correlation till converges to the error level characterized 
by noise in experimental data. In case of $J=0.9$, $\Delta J$ decreases to a minimum value then start increasing 
and finally converges to a point larger than 
the minimum value. When $J$ increases to $2.0$, 
$\Delta J$ keeps increasing after reaching the minimum point and finally goes beyond the initial error and diverges.
Coupling strength plays a role of inverse temperature, a larger coupling strength gives equivalently low temperature in the model
and dynamics of the model becomes more difficult to predict due to stronger correlations. Consequently, approximations 
we made in dynamical method turns worse with stronger couplings and results to worse convergence of 
gradient descent learning and larger inference error.

\begin{figure}
\includegraphics[width=\fw\textwidth]{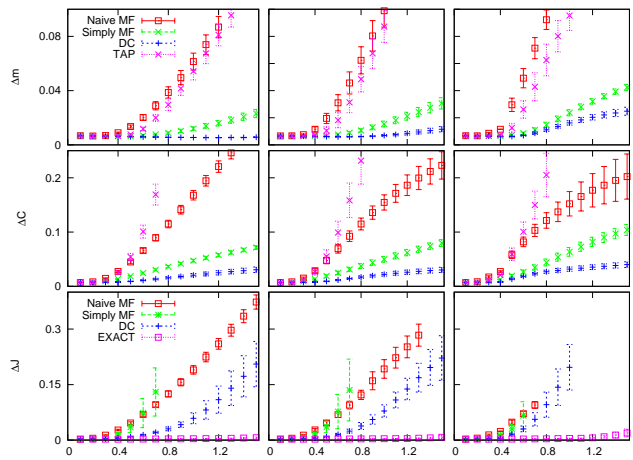}
\caption{ \label{fig:n100c5}
Difference of magnetizations $\Delta m$, one-step delayed correlations $\Delta C$ and inference error $\Delta J$ given by na\" ive
mean-field, TAP, simply mean-field and dynamical cavity method(DC) of Ising model 
with Gaussian distributed couplings with zero mean and variance $J$. $X$ axes denote coupling strength $J$. Size of the graph is $N=100$, average degree is $5$. 
Fraction of symmetric edges $P_s=0$ in left $3$ figures, $P_s=0.5$ in middle $3$ figures and in the right ones $P_s=1.0$.
Data are averaged over $10$ realizations. Some data of inference errors are not present because of un-convergence of gradient descent learning.
}
\end{figure}

We have compared convergence properties of gradient descent learning
with other approximations on the same network(data not shown), results show that 
with na\" ive mean-field, TAP and simply mean-field approximations, $\Delta J$ stops convergence
with couplings weaker than $0.9$ because they give even worse approximations than dynamical cavity method on larger couplings 
(see also fig.~\ref{fig:n100c5}).
However exact reconstruction does not have this converging problem as it does not use approximations so inference error only 
depends on quality of data(e.g. number of realizations used in computing experimental data), and is not a function of 
coupling strength.

To evaluate performance of approximations, especially how it is influenced by coupling strength, one method is 
to compare accuracy of approximations in direct problem, e.g., by applying approximations on graphs with fixed 
couplings and external fields to compute magnetization and correlation and compare result with experimental data. 
Error in correlations are characterized by eq.~(\ref{eq:dcj}) and now $\Delta C$ is not a function 
of time $t$ since couplings are fixed. Error of magnetization is defined in the same way by difference of magnetization given 
by approximations and that in the experimental data:
\begin{eqnarray}
  \Delta m=\frac{1}{\sqrt{N}}\sqrt{\sum_{<i>}(m_{i}^{J}-m_{i}^{data})^2}.
\end{eqnarray}

Result are plotted in fig.~\ref{fig:n100c5}, where graphs have same number of nodes and average connectivity but different degree of symmetry. Note that number of samplings used in computing experiment data is large enough, 
and noise in experimental data can be ignored compared with
error made in approximations, so a smaller $\Delta m$ and $\Delta J$ indicates a lower error made by the approximation.
Top and middle panels of fig.~\ref{fig:n100c5} show that in direct problem,
all approximations work better in asymmetric networks than in symmetric networks. This is because
in asymmetric networks, feedback correlations are not present so correlation ignored by approximations are fewer than in 
symmetric networks.
Among all method, dynamical method works the best in this direct problem, which indicates that Bethe approximations 
is more accurate than mean-field approximations in diluted networks.
More over, TAP method outperforms na\" ive mean-field only with weak couplings, because the expansion is carried out with 
weak couplings, and it is more difficult to find a solution in eq.~(\ref{eq:tapm}) with stronger couplings.

To evaluate performance of inference based on these approximations, we computed inference error $\Delta J$ 
given in eq. (\ref{eq:dj}) by different approximations on same 
set of graphs, results are plotted in bottom panel of fig.~\ref{fig:n100c5}.
TAP works badly and stops converging on networks with very weak coupling strength, 
so I did not plot the inference error of TAP. The missing points in the figure indicate that with value of that
coupling strength, gradient descent learning does not converge.
Figures show that all methods perform worse when degree of symmetry increases, as in the direct problems.
Dynamical cavity method converges with larger $J$ and gives smaller inference error at same $J$ 
than Na\" ive mean-field and simply mean-field. 
Simply mean-field outperforms na\" ive mean-field only with weak couplings and stops converging with weaker couplings. 
One interesting point is that, with degree of symmetry of network increases, 
the difference of $\Delta C$ and $\Delta J$ between dynamical cavity method and simply mean-field 
becomes larger while the difference between na\" ive mean-field and dynamical cavity becomes smaller. 

We saw from fig.~\ref{fig:con} and fig.~\ref{fig:n100c5} that, performances of 
dynamical cavity method is much worse than exact reconstruction. This is because due to computational expenses, networks 
we tested here are rather small, so loop effect can not be ignored on these networks (see e.g. \cite{Zhang_Chen_Physica_387_4411_2008}).
If we increase the sparsity of the graph, the error of dynamical cavity method should be decreased.
\begin{figure}
\includegraphics[width=\fw\textwidth]{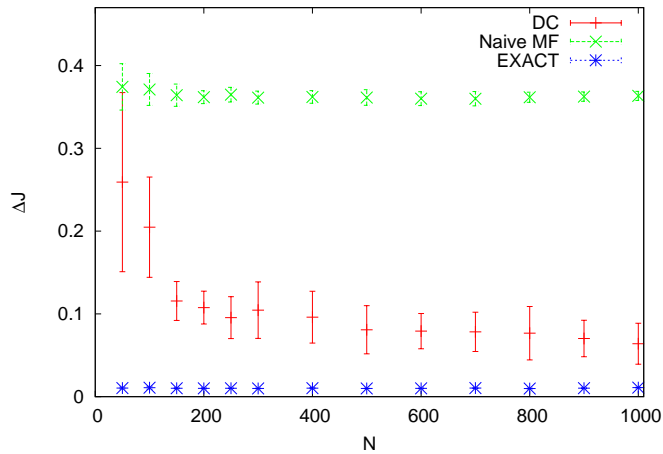}
\caption{ \label{fig:diffn}
Inference error of dynamical cavity inference(DC),  na\" ive mean-field and exact method for networks with different system size $N$.
Average connectivity of network is $5$, $P_s=0$, $J=1.5$ and data are averaged over $10$ realizations.  }
\end{figure}
To illustrate at this point, in fig.\ref{fig:diffn} we plot inference errors of dynamical cavity method for networks with 
same average connectivity but different system sizes. The figure shows that 
with $N$ increasing, inference error approaches the error of exact inference. We expect that with $N\to\infty$, 
inference error of dynamical cavity method gives same result to that of exact reconstruction in fully asymmetric networks.  
Note that in contrast to dynamical cavity method, mean-field method 
does not benefits from increasing system size because mean-field approximation does not become better with sparsity increases,
and as a consequence, performance between dynamical cavity method and mean-field methods becomes larger for sparse networks with larger
system sizes.

\section{conclusion and discussion}\label{sec:discuss}
In this paper we presented how to use dynamical cavity method to infer kinetic Ising model, and compared the performance on sparse graphs with existing mean-field methods.
Results show that in sparse graphs dynamical cavity method works better than other mean-field approximations and performance increases with sparsity of graph increases.
In computing cavity marginals, we use the simplest one-time approximation to simplify the cavity equations, and statistical quantities at time $t+1$ are computed from quantities 
at time $t$ and time $t-1$, which is different from mean-field approximations which do predictions by using only quantities at time $t$. We believe that dynamical cavity inference could 
benefit a lot from developing more efficient approximations in computing cavity equations to collect correlations between more time steps.
The networks discussed in this paper have known topology. It is difficult to apply dynamical cavity method to networks 
with unknown structure since it is so expensive to run cavity equations on fully-connected networks.
For inference problems which asks to reconstruct both network topology and coupling strengths, one can use
another method e.g. simply mean-field to reconstruct the structure then refine the couplings by cavity method.

\section{Acknowledgements}
The author would like to thank Abolfazl Ramezanpour and Riccardo Zecchina for discussing.
\input{fcising.bbl}

\begin{thebibliography}{10}%
\makeatletter
\providecommand \@ifxundefined [1]{%
 \ifx #1\undefined \expandafter \@firstoftwo
 \else \expandafter \@secondoftwo
\fi
}%
\providecommand \@ifnum [1]{%
 \ifnum #1\expandafter \@firstoftwo
 \else \expandafter \@secondoftwo
\fi
}%
\providecommand \enquote [1]{``#1''}%
\providecommand \bibnamefont  [1]{#1}%
\providecommand \bibfnamefont [1]{#1}%
\providecommand \citenamefont [1]{#1}%
\providecommand\href[0]{\@sanitize\@href}%
\providecommand\@href[1]{\endgroup\@@startlink{#1}\endgroup\@@href}%
\providecommand\@@href[1]{#1\@@endlink}%
\providecommand \@sanitize [0]{\begingroup\catcode`\&12\catcode`\#12\relax}%
\@ifxundefined \pdfoutput {\@firstoftwo}{%
 \@ifnum{\z@=\pdfoutput}{\@firstoftwo}{\@secondoftwo}%
}{%
 \providecommand\@@startlink[1]{\leavevmode\special{html:<a href="#1">}}%
 \providecommand\@@endlink[0]{\special{html:</a>}}%
}{%
 \providecommand\@@startlink[1]{%
  \leavevmode
  \pdfstartlink
   attr{/Border[0 0 1 ]/H/I/C[0 1 1]}%
   user{/Subtype/Link/A<</Type/Action/S/URI/URI(#1)>>}%
  \relax
 }%
 \providecommand\@@endlink[0]{\pdfendlink}%
}%
\providecommand \url  [0]{\begingroup\@sanitize \@url }%
\providecommand \@url [1]{\endgroup\@href {#1}{\urlprefix}}%
\providecommand \urlprefix [0]{URL }%
\providecommand \Eprint[0]{\href }%
\@ifxundefined \urlstyle {%
  \providecommand \doi [1]{doi:\discretionary{}{}{}#1}%
}{%
  \providecommand \doi [0]{doi:\discretionary{}{}{}\begingroup
  \urlstyle{rm}\Url }%
}%
\providecommand \doibase [0]{http://dx.doi.org/}%
\providecommand \Doi[1]{\href{\doibase#1}}%
\providecommand \bibAnnote [3]{%
  \BibitemShut{#1}%
  \begin{quotation}\noindent
    \textsc{Key:}\ #2\\\textsc{Annotation:}\ #3%
  \end{quotation}%
}%
\providecommand \bibAnnoteFile [2]{%
  \IfFileExists{#2}{\bibAnnote {#1} {#2} {\input{#2}}}{}%
}%
\providecommand \typeout [0]{\immediate \write \m@ne }%
\providecommand \selectlanguage [0]{\@gobble}%
\providecommand \bibinfo [0]{\@secondoftwo}%
\providecommand \bibfield [0]{\@secondoftwo}%
\providecommand \translation [1]{[#1]}%
\providecommand \BibitemOpen[0]{}%
\providecommand \bibitemStop [0]{}%
\providecommand \bibitemNoStop [0]{.\EOS\space}%
\providecommand \EOS [0]{\spacefactor3000\relax}%
\providecommand \BibitemShut [1]{\csname bibitem#1\endcsname}%
\bibitem{Cocco_etal_PNAS_206_14058_2009}%
  \BibitemOpen
  \bibfield{author}{%
  \bibinfo {author} {\bibfnamefont{S.}~\bibnamefont{Cocco}}, \bibinfo {author}
  {\bibfnamefont{S.}~\bibnamefont{Leibler}},\ and\ \bibinfo {author}
  {\bibfnamefont{R.}~\bibnamefont{Monasson}},\ }%
  \bibfield{journal}{%
  \bibinfo {journal} {PNAS}\ }%
  \textbf{\bibinfo {volume} {206}},\ \bibinfo {pages} {14058} (\bibinfo {year}
  {2009})%
  \bibAnnoteFile{NoStop}{Cocco_etal_PNAS_206_14058_2009}%
\bibitem{Weigt_etal_PNAS_2009}%
  \BibitemOpen
  \bibfield{author}{%
  \bibinfo {author} {\bibfnamefont{M.}~\bibnamefont{Weigt}}, \bibinfo {author}
  {\bibfnamefont{R.~A.}\ \bibnamefont{White}}, \bibinfo {author}
  {\bibfnamefont{H.}~\bibnamefont{Szurmant}}, \bibinfo {author}
  {\bibfnamefont{J.~A.}\ \bibnamefont{Hoch}},\ and\ \bibinfo {author}
  {\bibfnamefont{T.}~\bibnamefont{Hwa}},\ }%
  \bibfield{journal}{%
  \bibinfo {journal} {PNAS}\ }%
  \textbf{\bibinfo {volume} {106}},\ \bibinfo {pages} {67} (\bibinfo {year}
  {2009})%
  \bibAnnoteFile{NoStop}{Weigt_etal_PNAS_2009}%
\bibitem{Ackley1985147}%
  \BibitemOpen
  \bibfield{author}{%
  \bibinfo {author} {\bibfnamefont{D.~H.}\ \bibnamefont{Ackley}}, \bibinfo
  {author} {\bibfnamefont{G.~E.}\ \bibnamefont{Hinton}},\ and\ \bibinfo
  {author} {\bibfnamefont{T.~J.}\ \bibnamefont{Sejnowski}},\ }%
  \bibfield{journal}{%
  \Doi{10.1016/S0364-0213(85)80012-4}{\bibinfo {journal} {Cognitive Science}}\
  }%
  \textbf{\bibinfo {volume} {9}},\ \bibinfo {pages} {147 } (\bibinfo {year}
  {1985}),\ ISSN \bibinfo {issn} {0364-0213},\
  \url{http://www.sciencedirect.com/science/article/pii/S0364021385800124}%
  \bibAnnoteFile{NoStop}{Ackley1985147}%
\bibitem{Roudi_etal_PRE_79_051915}%
  \BibitemOpen
  \bibfield{author}{%
  \bibinfo {author} {\bibfnamefont{Y.}~\bibnamefont{Roudi}}, \bibinfo {author}
  {\bibfnamefont{J.}~\bibnamefont{Tyrcha}},\ and\ \bibinfo {author}
  {\bibfnamefont{J.}~\bibnamefont{Hertz}},\ }%
  \bibfield{journal}{%
  \bibinfo {journal} {Phys. Rev. E}\ }%
  \textbf{\bibinfo {volume} {79}},\ \bibinfo {pages} {051915} (\bibinfo {year}
  {2009})%
  \bibAnnoteFile{NoStop}{Roudi_etal_PRE_79_051915}%
\bibitem{Schneidman_etal_2006_nature}%
  \BibitemOpen
  \bibfield{author}{%
  \bibinfo {author} {\bibfnamefont{E.}~\bibnamefont{Schneidman}}, \bibinfo
  {author} {\bibfnamefont{M.~J.}\ \bibnamefont{Berry}}, \bibinfo {author}
  {\bibfnamefont{R.}~\bibnamefont{Segev}},\ and\ \bibinfo {author}
  {\bibfnamefont{W.}~\bibnamefont{Bialek}},\ }%
  \bibfield{journal}{%
  \bibinfo {journal} {Nature}\ }%
  \textbf{\bibinfo {volume} {440}},\ \bibinfo {pages} {1007} (\bibinfo {year}
  {2006})%
  \bibAnnoteFile{NoStop}{Schneidman_etal_2006_nature}%
\bibitem{Roudi_Hertz_PRL_106_048702}%
  \BibitemOpen
  \bibfield{author}{%
  \bibinfo {author} {\bibfnamefont{Y.}~\bibnamefont{Roudi}}\ and\ \bibinfo
  {author} {\bibfnamefont{J.}~\bibnamefont{Hertz}},\ }%
  \bibfield{journal}{%
  \Doi{10.1103/PhysRevLett.106.048702}{\bibinfo {journal} {Phys. Rev. Lett.}}\
  }%
  \textbf{\bibinfo {volume} {106}},\ \bibinfo {pages} {048702} (\bibinfo
  {month} {Jan}\ \bibinfo {year} {2011})%
  \bibAnnoteFile{NoStop}{Roudi_Hertz_PRL_106_048702}%
\bibitem{Mezard_Sakellariou_jstat_2011}%
  \BibitemOpen
  \bibfield{author}{%
  \bibinfo {author} {\bibfnamefont{M.}~\bibnamefont{Mezard}}\ and\ \bibinfo
  {author} {\bibfnamefont{J.}~\bibnamefont{Sakellariou}},\ }%
  \bibfield{journal}{%
  \bibinfo {journal} {J. Stat. Mech.}\ }%
  \textbf{\bibinfo {volume} {L07001}} (\bibinfo {year} {2011})%
  \bibAnnoteFile{NoStop}{Mezard_Sakellariou_jstat_2011}%
\bibitem{Sakellariou_etal_arxiv_2011}%
  \BibitemOpen
  \bibfield{author}{%
  \bibinfo {author} {\bibfnamefont{J.}~\bibnamefont{Sakellariou}}, \bibinfo
  {author} {\bibfnamefont{Y.}~\bibnamefont{Roudi}}, \bibinfo {author}
  {\bibfnamefont{M.}~\bibnamefont{Mezard}},\ and\ \bibinfo {author}
  {\bibfnamefont{J.}~\bibnamefont{Hertz}},\ }%
  \bibfield{journal}{%
  \bibinfo {journal} {arXiv}\ }%
  \textbf{\bibinfo {volume} {1106.0452}} (\bibinfo {year} {2011})%
  \bibAnnoteFile{NoStop}{Sakellariou_etal_arxiv_2011}%
\bibitem{Braunstein_etal_PRE_2011}%
  \BibitemOpen
  \bibfield{author}{%
  \bibinfo {author} {\bibfnamefont{A.}~\bibnamefont{Braunstein}}, \bibinfo
  {author} {\bibfnamefont{A.}~\bibnamefont{Ramezanpour}}, \bibinfo {author}
  {\bibfnamefont{R.}~\bibnamefont{Zecchina}},\ and\ \bibinfo {author}
  {\bibfnamefont{P.}~\bibnamefont{Zhang}},\ }%
  \bibfield{journal}{%
  \bibinfo {journal} {Phys. Rev. E}\ }%
  \textbf{\bibinfo {volume} {83}},\ \bibinfo {pages} {056114} (\bibinfo {year}
  {2011})%
  \bibAnnoteFile{NoStop}{Braunstein_etal_PRE_2011}%
\bibitem{Zeng_etal_PRE_2011}%
  \BibitemOpen
  \bibfield{author}{%
  \bibinfo {author} {\bibfnamefont{H.-L.}\ \bibnamefont{Zeng}}, \bibinfo
  {author} {\bibfnamefont{E.}~\bibnamefont{Aurell}}, \bibinfo {author}
  {\bibfnamefont{M.}~\bibnamefont{Alava}},\ and\ \bibinfo {author}
  {\bibfnamefont{H.}~\bibnamefont{Mahmoudi}},\ }%
  \bibfield{journal}{%
  \Doi{10.1103/PhysRevE.83.041135}{\bibinfo {journal} {Phys. Rev. E}}\ }%
  \textbf{\bibinfo {volume} {83}},\ \bibinfo {pages} {041135} (\bibinfo {month}
  {Apr}\ \bibinfo {year} {2011})%
  \bibAnnoteFile{NoStop}{Zeng_etal_PRE_2011}%
\bibitem{Coolen:arxiv:2000b}%
  \BibitemOpen
  \bibfield{author}{%
  \bibinfo {author} {\bibfnamefont{A.~C.~C.}\ \bibnamefont{Coolen}},\ }%
  \bibfield{journal}{%
  \bibinfo {journal} {arXiv:}\ }%
  \textbf{\bibinfo {volume} {cond-mat:}},\ \bibinfo {pages} {0006011} (\bibinfo
  {year} {2000}),\ \url{http://arxiv.org/abs/cond-mat/0006011v1}%
  \bibAnnoteFile{NoStop}{Coolen:arxiv:2000b}%
\bibitem{Hatchett_etal_JPA_37_6201_2004}%
  \BibitemOpen
  \bibfield{author}{%
  \bibinfo {author} {\bibfnamefont{J.~P.~L.}\ \bibnamefont{Hatchett}}, \bibinfo
  {author} {\bibfnamefont{B.}~\bibnamefont{Wemmenhove}}, \bibinfo {author}
  {\bibfnamefont{I.~P.}\ \bibnamefont{Castillo}}, \bibinfo {author}
  {\bibfnamefont{T.}~\bibnamefont{Nikoletopoulos}}, \bibinfo {author}
  {\bibfnamefont{N.~S.}\ \bibnamefont{Skantzos}},\ and\ \bibinfo {author}
  {\bibfnamefont{A.~C.~C.}\ \bibnamefont{Coolen}},\ }%
  \bibfield{journal}{%
  \bibinfo {journal} {J. Phys. A: Math. Gen.}\ }%
  \textbf{\bibinfo {volume} {37}},\ \bibinfo {pages} {6201} (\bibinfo {year}
  {2004})%
  \bibAnnoteFile{NoStop}{Hatchett_etal_JPA_37_6201_2004}%
\bibitem{Coolen_Sherrington_PRB_1996}%
  \BibitemOpen
  \bibfield{author}{%
  \bibinfo {author} {\bibfnamefont{A.~C.~C.}\ \bibnamefont{Coolen}}, \bibinfo
  {author} {\bibfnamefont{S.~N.}\ \bibnamefont{Laughton}},\ and\ \bibinfo
  {author} {\bibfnamefont{D.}~\bibnamefont{Sherrington}},\ }%
  \bibfield{journal}{%
  \Doi{10.1103/PhysRevB.53.8184}{\bibinfo {journal} {Phys. Rev. B}}\ }%
  \textbf{\bibinfo {volume} {53}},\ \bibinfo {pages} {8184} (\bibinfo {month}
  {Apr}\ \bibinfo {year} {1996}),\
  \url{http://link.aps.org/doi/10.1103/PhysRevB.53.8184}%
  \bibAnnoteFile{NoStop}{Coolen_Sherrington_PRB_1996}%
\bibitem{Sommers_PRL_1987}%
  \BibitemOpen
  \bibfield{author}{%
  \bibinfo {author} {\bibfnamefont{H.-J.}\ \bibnamefont{Sommers}},\ }%
  \bibfield{journal}{%
  \Doi{10.1103/PhysRevLett.58.1268}{\bibinfo {journal} {Phys. Rev. Lett.}}\ }%
  \textbf{\bibinfo {volume} {58}},\ \bibinfo {pages} {1268} (\bibinfo {month}
  {Mar}\ \bibinfo {year} {1987}),\
  \url{http://link.aps.org/doi/10.1103/PhysRevLett.58.1268}%
  \bibAnnoteFile{NoStop}{Sommers_PRL_1987}%
\bibitem{Neri_Bolle_JSM_2009}%
  \BibitemOpen
  \bibfield{author}{%
  \bibinfo {author} {\bibfnamefont{I.}~\bibnamefont{Neri}}\ and\ \bibinfo
  {author} {\bibfnamefont{D.}~\bibnamefont{Bolle}},\ }%
  \bibfield{journal}{%
  \bibinfo {journal} {J. Stat. Mech.},\ \bibinfo {pages} {P08009}}%
   (\bibinfo {year} {2009})%
  \bibAnnoteFile{NoStop}{Neri_Bolle_JSM_2009}%
\bibitem{Aurell_Mahmoudi_JSM_2011}%
  \BibitemOpen
  \bibfield{author}{%
  \bibinfo {author} {\bibfnamefont{E.}~\bibnamefont{Aurell}}\ and\ \bibinfo
  {author} {\bibfnamefont{H.}~\bibnamefont{Mahmoudi}},\ }%
  \bibfield{journal}{%
  \bibinfo {journal} {J. Stat. Mech.}\ }%
  \textbf{\bibinfo {volume} {P04014}} (\bibinfo {year} {2011})%
  \bibAnnoteFile{NoStop}{Aurell_Mahmoudi_JSM_2011}%
\bibitem{Zhang_Chen_Physica_387_4411_2008}%
  \BibitemOpen
  \bibfield{author}{%
  \bibinfo {author} {\bibfnamefont{P.}~\bibnamefont{Zhang}}\ and\ \bibinfo
  {author} {\bibfnamefont{Y.}~\bibnamefont{Chen}},\ }%
  \bibfield{journal}{%
  \bibinfo {journal} {Physica A}\ }%
  \textbf{\bibinfo {volume} {387}},\ \bibinfo {pages} {4441} (\bibinfo {year}
  {2008})%
  \bibAnnoteFile{NoStop}{Zhang_Chen_Physica_387_4411_2008}%
\end{thebibliography}%
\end{document}